\documentstyle[12pt,epsf]{article}
\input{psfig.sty}
\newcommand{\sptwo}{1.6}

\newcommand{\doublespace}{\edef\baselinestretch{\sptwo}\Large\normalsize}

\title{ The effect of Ramsauer type transmission resonances on the conductance
modulation of spin interferometers}
\author{M. Cahay\\
Department of Electrical and Computer Engineering and Computer Science \\
University of Cincinnati, Cincinnati, Ohio 45221\\
\\
S. Bandyopadhyay\\
Department of Electrical Engineering \\
Virginia Commonwealth University, Richmond, Virginia 23284}
\date{}

\begin{document}

\maketitle

\doublespace

\begin{abstract}
\noindent
We use a mean field approach to study the conductance modulation of gate
controlled semiconductor spin interferometers  based on the Rashba spin-orbit
coupling effect. The conductance modulation is found to be mostly due to Ramsauer type
transmission resonances  rather than the Rashba effect in typical structures.
This is because of significant reflections at the interferometer's
contacts due to large potential barriers and effective mass mismatch between the contact 
material and the semiconductor. Thus, unless particular care is taken to
eliminate these reflections, any observed conductance modulation of spin
interferometers may have its origin in the Ramsauer resonances (which is unrelated to spin)
rather than the Rashba effect.


\end{abstract}

\bigskip

\noindent {\bf PACS}: 72.25.Dc, 72.25.Mk, 73.21.Hb, 85.35.Ds

\pagebreak
In a seminal paper, Datta and Das \cite{datta} proposed a quasi one-dimensional 
gate controlled ballistic spin interferometer which is an analog of the standard 
electro-optic light modulator. Their device consists of a  one-dimensional 
semiconductor 
channel with ferromagnetic source and drain contacts (Fig.1). Electrons are 
injected with a definite spin orientation into the channel from the source,  
then controllably precessed in the channel with a gate voltage which  
varies the Rashba interaction in the channel \cite{rashba}, and finally sensed 
at the drain. At the drain end, the electron's transmission probability depends 
on the relative 
alignment of its spin with  the drain's (fixed) magnetization.  By controlling 
the angle of spin precession in the channel with a gate voltage, 
one can modulate the relative spin alignment at the drain end, and hence control 
the 
source-to-drain  current (or conductance). This is the basis of 
the gate controlled spin interferometer.

The work of Datta and Das motivated vigorous research in the field of 
spintronics. However, to our knowledge,  there has never been a complete 
calculation of the spin interferometer's conductance as a function of the gate 
voltage in realistic structures. This would require solving the Schr\"odinger 
equation (with the Rashba 
spin-orbit coupling 
term in the Hamiltonian) self consistently with the Poisson equation to extract 
the transmission probability 
of incident electrons through the channel, and then using the Landauer formula 
to evaluate the conductance. 
In this Letter, we provide a mean-field approach to this problem.

In any realistic spin interferometer structure, varying 
the gate 
voltage will inevitably move the Fermi level in the interferometer's channel up 
or down relative to the conduction band edge. We must therefore consider the 
following physical mechanism as another potential source
of conductance modulation of the spin interferometer.  Referring to Fig.2, if we 
neglect the Rashba effect 
momentarily, then it is well known that the transmission
through the semiconducting channel of the interferometer (barrier region) should 
peak each time the Fermi level
lines up with the resonant energy levels above the barrier between the two 
contacts \cite{ohanian}. These levels are given by
\begin{equation}
E_n = V_0 + \frac{ n^2 {\hbar}^2 {\pi}^2 }{ {2 m^* L^2}},
\end{equation}
where n is an integer, $V_0$ is the height of the potential barrier between the 
ferromagnetic
contacts  and the semiconducting channel (including the effects of quantum 
confinement in the directions transverse to the channel axis),
$m^*$ is the effective mass in the semiconductor and L is the length of the 
channel. As the gate voltage is varied, the Fermi level sweeps through the 
resonant levels causing the conductance to oscillate. This is the Ramsauer 
effect. In this Letter, we will show that the Ramsauer effect completely masks 
the Rashba effect and becomes the primary source of any conductance modulation 
of the spin interferometer shown in Fig.1.

The quasi one-dimensional spin interferometer is described by the single 
particle effective-mass Hamiltonian \cite{moroz}
\begin{equation}
{\cal H} = {{1}\over{2 m^*}} \left ( {\vec p} + e{\vec A} \right )^2 
+ V(y) + V(z) 
-  (g/2) \mu_B {\vec B} \cdot {\vec \sigma} 
+  \frac{{\alpha}_R}{ \hbar} \hat{y}\cdot \left [ {\vec \sigma} \times ( {\vec 
p} + e {\vec A} ) 
\right ]
\end{equation}
where $\hat{y}$ is the unit vector normal to the heterostructure interface in 
Fig.1 and ${\vec A}$ is 
the vector potential due to the axial magnetic field ${\vec B}$ along the 
channel caused by the ferromagnetic contacts (this magnetic field was summarily 
ignored in all previous work, but has important consequences). 
The Rashba coupling strength ${\alpha}_R$ varies with the applied potential on 
the gate.  We will assume that the 
confining potential $V(z)$ along the z-direction  is parabolic.

The choice of the Landau gauge ${\vec A}$ = (0, -Bz, 0)  allows us to decouple 
the y-component of the Hamiltonian in (2) from the x-z component.
Furthermore, since this Hamiltonian is translationally invariant in the 
x-direction, the wavevector $k_x$ is a good quantum number and the eigenstates 
are plane waves 
traveling in the x-direction.  The two-dimensional Hamiltonian in the plane of 
the channel (x-z plane) is then given by
\begin{equation}
H_{xz} = {{p_z^2}\over{2 m^*}} + \Delta E_c + {{1}\over{2}}m^* \left ( 
\omega_0^2 + 
\omega_c^2 \right ) z^2 + {{\hbar^2 k_x^2}\over{2 m^*}} +{{\hbar^2 k_R 
k_x}\over{m^*}} \sigma_z - ( g^* /2) \mu_B B {\sigma}_x  - {{\hbar k_R 
p_z}\over{m^*}} \sigma_x
\label{Hamiltonian}
\end{equation}
where $\omega_0$ is the curvature of the confining potential in the z-direction,
$\omega_c$ = $eB/m^*$, ${\mu}_B$ is the Bohr magneton, $g^*$ is the 
magnitude of the Land\'{e} factor of the electron in the channel, $k_R = m^* 
\alpha_R/\hbar^2$, 
and $\Delta E_c$ is the  potential barrier between the ferromagnet and 
semiconductor. We assume that $\Delta E_c$ includes the effects of the quantum 
confinement in the y-direction.

We now derive the energy dispersion relations in the channel of the 
interferometer from Equation (\ref{Hamiltonian}).  The first five
terms of the Hamiltonian in Eq.(\ref{Hamiltonian}) 
yield shifted parabolic subbands with dispersion relations:
\begin{equation}
E_{n, \uparrow} = ( n + 1/2) \hbar \omega + \Delta E_c
+ {{\hbar^2 k_x^2}\over{2 m^*}} + {{\hbar^2 k_R k_x}\over{m^*}}, ~~~
E_{n, \downarrow} = ( n + 1/2) \hbar \omega + \Delta E_c
+ {{\hbar^2 k_x^2}\over{2 m^*}} - {{\hbar^2 k_R k_x}\over{m^*}},
\end{equation}
where $\omega = 
\sqrt{ \omega_0^2 + \omega_c^2 }$. In Eq.(4), the $\uparrow$ and $\downarrow$ 
arrows indicate +z and -z 
polarized spins (eigenstates of the $\sigma_z$ operator) which are split by the 
Rashba effect. These are unperturbed 
subbands with definite 
spin quantizations axes along +z and -z directions. Their dispersion relations 
are shown as dashed lines in Fig. 1.

The sixth and seventh terms in Eq.(3) induce a perturbation and mixing between 
the 
unperturbed subbands (+z- and -z-polarized spins). 
The sixth term originates from the magnetic field due to the ferromagnetic 
contacts and the seventh originates 
from the Rashba effect itself. The sixth term was ignored and the seventh was 
assumed to be negligibly small in ref. \cite{datta}.  The ratio of the 
sixth and seventh term can be shown to be of the order of 
10$^4$ - 10$^6$ for typical values of the relevant parameters. 
Therefore, we can neglect the seventh term in comparison with the sixth term.

To obtain an analytical expression for the dispersion relation 
corresponding to the first five terms in the Hamiltonian in Eq.(3),  we 
derive the two-band dispersion relation in a truncated Hilbert space considering 
mixing between the 
two lowest unperturbed subband states (namely the +z and -z spin 
states).
Straightforward diagonalization of the Hamiltonian in Eq.(3) (minus the 
seventh term) in the basis of these two unperturbed states gives the following 
dispersion relations:
\begin{equation}
E_1 (k_x) = {{1}\over{2}} \hbar \omega + \Delta E_c + {{\hbar^2 k_x^2}\over{2 
m^*}}
- \sqrt{ \left ({{\hbar^2 k_R k_x}\over{m^*}} \right )^2 + \left (
\frac{ g^* \mu_B B}{2} \right )^2 },
\label{dispersion1}
\end{equation}
\begin{equation}
E_2 (k_x) = {{1}\over{2}} \hbar \omega + \Delta E_c + {{\hbar^2 k_x^2}\over{2 
m^*}}
+ \sqrt{ \left ( {{\hbar^2 k_R k_x}\over{m^*}} \right )^2 + \left (
\frac{ g^* \mu_B B}{2} \right )^2 },
\label{dispersion2}
\end{equation}
where the indices 1 and 2 refer to the lower and upper subbands. Their 
dispersion
relations are plotted schematically as solid lines in Fig.1.

One can see from Fig.1 that the magnetic field caused by the ferromagnetic 
contacts couples the two unperturbed 
subbands and changes their dispersion relation, lifting the degeneracy at $k_x$ 
= 0. While the unperturbed
bands are shifted parabolas with single minima at $k_x = \pm k_R$ \cite{datta}, 
the perturbed bands (in the presence of a magnetic field) are not parabolic and 
are symmetric 
about the energy axis. One of them has a single minimum at $k_x$ = 0, and the 
other has double 
minima at 
$k_x = \pm k_R \sqrt{1 + ( g^* \mu_B B / \delta_{R})^2}$,
where $\delta_{R}$ = ${\hbar}^2 k_R^2/2m^*$. The magnetic field not only has 
a profound influence on the dispersion relations, but it also causes {\it spin 
mixing}, meaning that the perturbed subbands no longer have definite 
spin quantization axes (spin quantization becomes wavevector dependent). 
Furthermore, energy degenerate states in the two perturbed subbands no longer 
have orthogonal spins. Therefore, elastic scattering between them is possible without 
a complete spin flip. 

The energy dispersion relations also show that, in a semiconductor where the 
Zeeman splitting energy can be made comparable to the Rashba spin-splitting 
energy ${\delta}_R$, the difference $\Delta k_x$ between
the wavevectors in the two subbands at any energy is {\it not} 
independent of energy. Since this difference is proportional to the angle by 
which the spin precesses in the channel \cite{datta}, the angle of spin 
precession is {\it no longer} independent of electron energy. Thus different 
electrons that are injected from the contact with different energies (at finite 
temperature and bias) will undergo different degrees of spin precession, 
and the conductance modulation will not survive ensemble averaging over 
a broad spectrum of electron energy at elevated temperatures and bias. 

From Equations (\ref{dispersion1} - \ref{dispersion2}), we find that  for an 
electron incident with total energy E, the corresponding wavevectors in the two 
subbands are given by
\begin{equation}
k_{x \pm} = \frac{1}{\hbar} \sqrt{ 2m^*  ( \frac{ B \pm \sqrt{B^2 - 4C}  }{2}) 
},
\end{equation}
where
\begin{equation}
B = 2 (E - \frac{ \hbar \omega}{2} - \Delta E_c ) + 4 {\delta}_{R}, ~~~
C = ( E - \frac{ \hbar \omega}{2} - \Delta E_c )^2 - {\beta}^2,
\end{equation}
with $\beta$ = $ g^* \mu_B B/2 $.

In Eq.(8), the upper and lower sign corresponds to the 
lower and upper subbands and are referred to hereafter as $k_{x,1}$ and 
$k_{x,2}$, respectively.
The corresponding eigenvectors for the two subbands are respectively
\begin{eqnarray}
\left [ \begin{array}{c}
             C_1 (k_{x,1})\\
             {C^{'}}_1 (k_{x,1})\\
             \end{array}   \right ]
& = &
 \left [ \begin{array}{c}
- \alpha (k_{x,1})/\gamma (k_{x,1})\\
\beta/\gamma (k_{x,1}) \\
\end{array} \right ] \nonumber \\
\left [ \begin{array}{c}
             C_2 (k_{x,2})\\
             {C^{'}}_2 (k_{x,2})\\
             \end{array}   \right ]
& = &
 \left [ \begin{array}{c}
 \beta/\gamma (k_{x,2}) \\
\alpha (k_{x,2}) /\gamma (k_{x,2})\\
\end{array} \right ]
\end{eqnarray}
where the quantities $\alpha$ and $\gamma$ are function of $k_{x}$ and are given 
by
\begin{equation}
\alpha (k_{x})  =  {{\hbar^2 k_R k_x}\over{m^*}} + \sqrt{ \left ( {{\hbar^2 
k_R 
k_x}\over{m^*}} \right )^2 + {\beta}^2 }, ~~~
\gamma ( k_x ) =  \sqrt{ \alpha^2 + \beta^2}.
\end{equation}

Note that the eigenspinors given by Eq.(9) are not 
+z-polarized state $\left 
[ \begin{array}{cc}
1 & 0 
\end{array} \right ]^{\dagger}
$,  or -z-polarized state $\left [ \begin{array}{cc}
0 & 1 
\end{array} \right ]^{\dagger} 
$ if the magnetic field $B \neq 0$. Thus, the magnetic field mixes spins
and the +z or -z polarized states are no longer eigenstates in the channel.
Equations (9) also show that the spin quantization (eigenspinor) in any subband 
is not fixed and strongly depends on the wavevector $k_x$. Thus, an electron 
entering 
the semiconductor channel from the left ferromagnetic contact with +x-polarized 
spin, will not couple {\it 
equally} to +z and -z states. The relative coupling will depend on the 
electron's energy. 

We model the ferromagnetic contacts by 
the Stoner-Wohlfarth model. The 
spin-up (majority carriers) and spin-down (minority carriers) band 
energies are offset by an exchange splitting energy $\Delta$ (Fig.2).

Next, we calculate the 
total transmission coefficient through the spin interferometer for an electron 
entering from the left ferromagnetic  contact (region I) and 
exiting at the right ferromagnetic contact (region III). 
A rigorous treatment of this problem would require an accurate modeling of the 
three- to one-dimensional transition between the bulk ferromagnetic contacts 
(regions I and III) and the
quantum wire semiconductor
channel (region II) \cite{kriman,frohne}. However, a one-dimensional transport 
model to 
calculate 
the transmission coefficient through the structure is known to be a very good 
approximation when the Fermi wave number in the ferromagnetic
contacts is much greater than the inverse of the transverse dimensions of the 
quantum wire \cite{grundler,raichev}. This is always the case with metallic 
contacts.

In region II ( $ 0 < x < L$), the
x-component of the wavefunction at a position $x$ along  
the channel is given by
\begin{eqnarray}
{\psi}_{II} (x) & = &
A_I
 \left [ \begin{array}{c}
             C_1(k_{x,1})\\
             C'_1 (k_{x,1})\\
             \end{array}   \right ]
             ^{i k_{x,1} x}
             +
A_{II}
 \left [ \begin{array}{c}
             C_1(-k_{x,1})\\
             C'_1 (-k_{x,1})\\
             \end{array}   \right ]
             e^{-i k_{x,1} x} \nonumber \\
& &   
             +           
A_{III}
 \left [ \begin{array}{c}
             C_2(k_{x,2})\\
             C'_2 (k_{x,2})\\
             \end{array}   \right ]
             e^{i k_{x,2} x}
             +
 A_{IV}
 \left [ \begin{array}{c}
             C_2(-k_{x,2})\\
             C'_2 (-k_{x,2})\\
             \end{array}   \right ]
             e^{-i k_{x,2} x} .
\label{wavefunction}
\end{eqnarray}

For a spin-up electron in the left ferromagnetic contact (region I; $x < 0$), 
the electron is spin polarized in the
$ \left [ \begin{array}{cc}
             1, 1
             \end{array}
             \right ]^{\dagger}  $
subband and the x-component of the wavefunction 
is given by
\begin{eqnarray}
{\psi}_I (x) & = &
 \frac{1}{\sqrt{2}} \left [ \begin{array}{c}
             1\\
             1\\
             \end{array}   \right ]
             e^{i {k_{x}}^u x}
             +
  \frac{ R_1}{\sqrt{2}} \left [ \begin{array}{c}
             1\\
             1\\
             \end{array}   \right ]
             e^{-i {k_{x}}^u x} 
             +
 \frac{R_2}{\sqrt{2}} \left [ \begin{array}{c}
              1\\
             -1\\
             \end{array}   \right ]
             e^{-i {k_{x}}^d  x}.
\end{eqnarray}
where $R_1$ is the reflection amplitude into the spin-up band and $R_2$ is the 
reflection amplitude in the spin-down band.

In region III $(x > L)$, the x-component of the 
wavefunction 
is given by
\begin{equation}
{\psi}_{III} (x)  = 
 \frac{T_1}{\sqrt{2}} \left [ \begin{array}{c}
             1\\
             1\\
             \end{array}   \right ]
             e^{i {k_{x}}^u  (x-L)}
             +
 \frac{T_2}{\sqrt{2}} \left [ \begin{array}{c}
             1\\
             -1\\
             \end{array}   \right ]
             e^{i {k_{x}}^d (x-L)}  .
\end{equation}
where $T_1$ and $T_2$ are the transmission amplitudes into the spin-up and 
spin-down bands. The wavevectors
\begin{equation}
k_x^u  = \frac{1}{ \hbar} \sqrt{2 m_0 E}, ~~~ k_x^d = \frac{1}{ \hbar} \sqrt{2 
m_0 (E - \Delta)},
\end{equation}
are the x components of the wavevectors in the spin-up and spin-down energy 
bands, respectively.

The eight unknowns ($R_1$,$R_2$,$T_1$,$T_2$,$A_{i} (i=I,II,III,IV))$ must
be found by enforcing continuity of the wavefunction and the quantity
$ \frac{1}{m^* (x) } (\frac{d \psi }{dx} + i k_R (x) {\sigma}_z \psi (x))$
at x = 0 and x = L. The latter condition insures continuity
of the current density.  This leads to a system of 8 coupled equations for the 
unknowns 
which must be solved to extract the transmission amplitudes $T_1$,$T_2$ in the 
spin-up and spin-down energy bands in the right ferromagnetic contact.

For the majority spin carriers, 
the linear response source-to-drain conductance 
of the spin interferometer at any 
temperature $T$ is given by  the Landauer formula
\begin{equation}
G_{\uparrow} = {{e^2}\over{4 h kT}} \int_0^{\infty} dE |T_{tot} (E)|^2 sech^2 
\left ( {{E - 
E_F}\over{2 kT}} \right ),
\label{conductance}
\end{equation}
where
\begin{equation}
|T_{tot} (E)|^2 = |T_{1} (E)|^2 + ({k_{x}}^d / {k_{x}}^u ) |T_{2} (E)|^2
\end{equation}
is the total transmission coefficient through the interferometer.

Similarly, the conductance of the minority spin carriers ($G_{\downarrow}$) is calculated after 
repeating the scattering problem for electrons incident from the minority spin band in the 
contacts.  Since the up- and down-spin states are orthogonal in the contacts, the total 
conductance of the spin interferometer is then given by $G_{\uparrow} $ + 
$G_{\downarrow} $.

We consider a spin interferometer consisting of 
a quasi one-dimensional InAs 
channel between two ferromagnetic contacts.  The electrostatic potential in the 
z-direction is 
assumed to be harmonic with $\hbar \omega$ = 10 meV in Eq.(4).
A Zeeman splitting energy of 0.34 meV is used in the semiconductor channel 
assuming a magnetic field B = 1 Tesla along the channel. 
This corresponds to a  $g^*$ factor of 3 and an electron effective mass $m^* = 
0.036 m_o$ which is typical of InAs-based channels \cite{datta}. The Fermi level 
$E_f$ and the exchange splitting energy $\Delta $ in the ferromagnetic contacts 
are set equal to 4.2 and 3.46 eV, respectively  \cite{mireles}.

The Rashba spin-orbit coupling strength ${\alpha}_R$ is 
typically derived from low-temperature magnetoresistance measurements 
(Shubnikov-de Haas 
oscillations) in 2DEG created at the interface of 
semiconductor heterostructures \cite{nitta}. To date, the largest reported 
experimental 
values of the Rashba spin-orbit coupling strength ${\alpha}_R$ has been found in 
InAs-based semiconductor 
heterojunctions. For a normal HEMT 
$In_{0.75} Al_{0.25} As/In_{0.75} Ga_{0.25} As $ heterojunction, Sato et al. 
have reported variation 
of ${\alpha}_R$ from 30- to 15 $\times 10^{-12}$ eVm when the external gate 
voltage is swept from 0 to -6 V (the total electron 
concentration in the 2DEG is found to be reduced from 5- to 4.5$\times 10^{11} 
/cm^2$ over the same range of bias). 
For a channel length of 0.1 $\mu m$, this corresponds to a variation of the
spin precession angle $\theta$ = $2 k_R L$ from about $\pi$ to $0.5 \pi$ over 
the same range 
of gate bias.

In the numerical results below, we calculated the conductance of a spin 
interferometer with a 0.2 $\mu m$ long channel
as a function of the gate voltage. Tuning the gate voltage varies the potential 
energy barrier $\Delta E_c$. Therefore, we have effectively calculated the 
interferometer's conductance as a function of $\Delta E_c$.
In our calculations, we vary $\Delta E_c$ over a range of 10 meV
which allows us to display several of the Ramsauer oscillations for
the selected separation between source and drain. 
The final energy $\Delta E_c$ is equal to the Fermi energy $E_f$. At that point, 
the 
Fermi energy lines up with the top of the potential barrier which corresponds to 
complete pinch-off 
of the channel when the carrier concentration falls to zero. 
Over that range of $\Delta E_c$, we simulated several
cases of Rashba spin-orbit coupling strength ${\alpha}_R$ variation with 
increasing $\Delta E_c$ (or increasing gate voltage): {\bf Case 1}:
${\alpha}_R$ stays constant and is equal to the largest experimental value 
reported to date (30$\times 10^{-12}$eVm), {\bf Case 2}: ${\alpha}_R$ varies from 30 $\times 10^{-12}$ 
eVm down to zero, and {\bf Case 3}: ${\alpha}_R$ varies from
zero to a maximum of 30 $ \times 10^{-12}$ eVm, which is the reverse of the 
previous case. A situation where ${\alpha}_R$ actually increases with reduction 
of the carrier concentration in the channel was reported for inverted 
InAlAs/InGaAs 
heterostructures by Schapers et al. \cite{schapers2}. Finally, we 
consider {\bf Case 4} where ${\alpha}_R$ is varied from 3 $\times 10^{-10}$ eVm 
(a tenfold improvement over the largest reported experimental result) down to 
zero. 
This last case corresponds to a variation of the spin precession angle 
$\theta$ from about $10 \pi$ to 0 over 
the range of $\Delta E_c $ considered.

The results of the conductance modulation are shown in Fig.3 for the four cases
described above at T = 2 K. This figure shows that there is very little change
between the different curves corresponding to cases 1 through 3
of the ${\alpha}_R$ dependence on $\Delta E_c$.
The location of the resonant energy levels was calculated using Eq.(1) and the
various quantum numbers $n$ characterizing the subbands being swept through the 
Fermi energy
are indicated in the figure. The gate voltage variation of the Rashba spin 
splitting energy
modifies slighty the shape and position of the resonant peaks due to  
electrostatic adjustment of
the potential barrier between the two ferromagnetic contacts.
Even for case 4, the amplitude of the conductance
oscillations are virtually unchanged and merely shifted along the $\Delta E_c$
axis compared to cases 1 through 3.

Referring to Fig.1, it can be seen that the energy dispersion relations are not 
parallel as the Fermi level approaches the bottom of the upper subband in the 
channel, 
i.e, near channel pinch-off (at larger value of $\Delta E_c$). Furthermore, the 
oscillations in conductance are more closely spaced as the quasi 1D channel 
approaches 
pinch-off. These two features make the conductance modulation near pinch-off 
more sensitive 
to temperature averaging. As illustrated in Fig.3, the conductance oscillations
are washed out completely for T = 10 K. This is only shown for Case 2 but 
similar degradation of the conductance modulation with temperature is found for 
all other cases considered above.

In conclusion,
we have demonstrated that the conductance modulation 
of typical electron spin interferometer structures may be primarily due to the 
Ramsauer effect 
\cite{ohanian} rather than the Rashba effect. The Ramsauer 
effect is caused by strong reflections at the contact-channel interfaces which 
are exacerbated by the large value of $\Delta E_c$ and the significant effective 
mass differences between the ferromagnetic contact material and the semiconductor. 
We have also found that the Ramsauer oscillations are accentuated when an insulating barrier is interposed at the 
ferromagnet-semiconductor interface \cite{eirashba} since it enhances multiple 
reflections in the channel.
Finally, we have studied the effect of 
elastic scattering in the channel using the scattering matrix technique of ref. 
\cite{cahay}. 
The details will be presented elsewhere, but a few elastic scatterers in the 
channel do not affect 
the results significantly.  Thus, unless the interferometer is well-designed to 
eliminate contact reflections, any experiment that purports to demonstrate the 
spin interferometer needs to pay careful attention 
to the actual origin of the oscillations.

\vskip .1in

M. C. dedicates this article to the memory of his father-in-law.
The authors acknowledge insightful discussions with S. Datta.
The work of  S. B. is supported by the 
National Science Foundation under grant ECS-0089893.

\pagebreak

\noindent {\bf Figure Captions}

\noindent {\bf Fig. 1}: A schematic of the electron spin interferometer
from ref. \cite{datta}.  The horizontal dashed line represents the quasi 
one-dimensional electron gas formed at the semiconductor interface between 
materials I and II.
The magnetization of the ferromagnetic contacts is assumed to be along the
+x-direction which results in a magnetic field along the x-direction.  
Also shown is a qualitative representation of the energy dispersion of the two 
perturbed 
(solid line) and unperturbed (broken line)  bands under the gate. The 
unperturbed bands are given by Equation (4) and the perturbed ones are
given by Equations (5) and (6) in the text. 

\vskip .2in
\noindent {\bf Fig. 2}: Energy band diagram across the electron spin 
interferometer.  We use a
Stoner-Wohlfarth model for the ferromagnetic contacts. $\Delta$ is the
exchange splitting energy in the contacts. $V_o$ is the height of the potential 
barrier between the energy
band bottoms of the semiconductor and the ferromagnetic electrodes. $V_o$ takes 
into
account the effects of the quantum confinement in the y- and z-directions.  Also 
shown as dashed
lines are the resonant energy states above $V_o$. Peaks in the conductance of 
the
electron spin interferometer are expected when the Fermi level in the contacts 
lines up with the
resonant states.

\vskip .2in
\noindent {\bf Fig. 3}: 
Conductance modulation of the electron spin
interferometer  (for T = 2 K) for different variations of the Rashba spin-orbit
coupling strength ${\alpha}_R$ with the energy barrier $\Delta E_c$. The Fermi 
energy
$E_f$ is designated in the figure. The different
${\alpha}_R$ vs. $\Delta E_c$ variations are labeled $\#$ 1 through $\# 4$
corresponding to cases 1 through 4 in the text.
The separation between the two ferromagnetic contacts is
 0.2 $\mu m$.  The confinement energy $\hbar \omega$ is 
10 meV. We have indicated the conductance peaks corresponding to different 
resonant energy
levels lining up with the Fermi level in the contacts. The curve labeled T = 10 
K 
represents the conductance modulation computed at a temperature of 10 K
when ${\alpha}_R$ varies from 30 $\times 10^{-12}$ eVm to 0 as the gate voltage 
is
varied.

\newpage
\
\vskip .2in
\begin{figure}[h]
\centerline{\psfig{figure=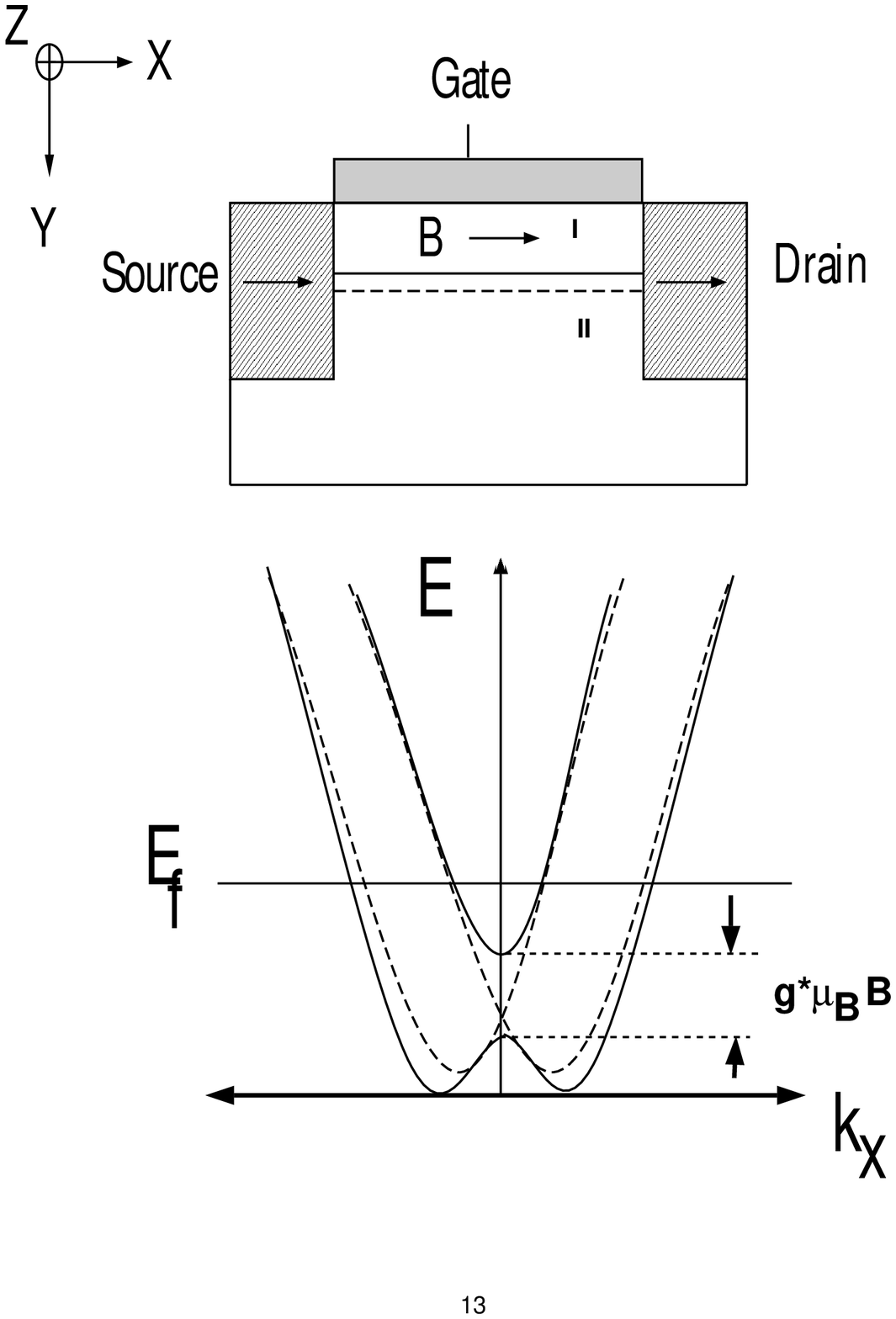,height=5.5in,width=6in}}
\end{figure}

\newpage
\
\vskip .2in
\begin{figure}[h]
\centerline{\psfig{figure=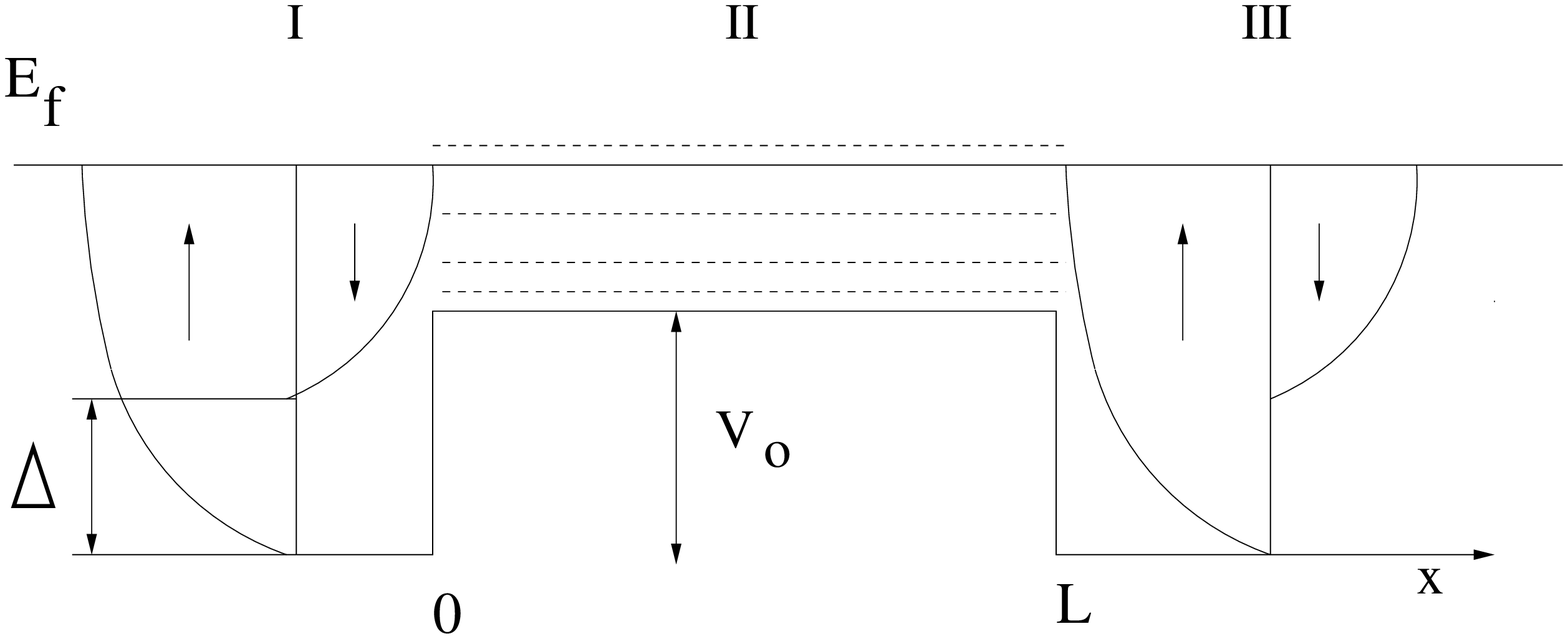,height=4.5in,width=5.5in}}
\end{figure}

\newpage
\
\vskip .2in
\begin{figure}[h]
\centerline{\psfig{figure=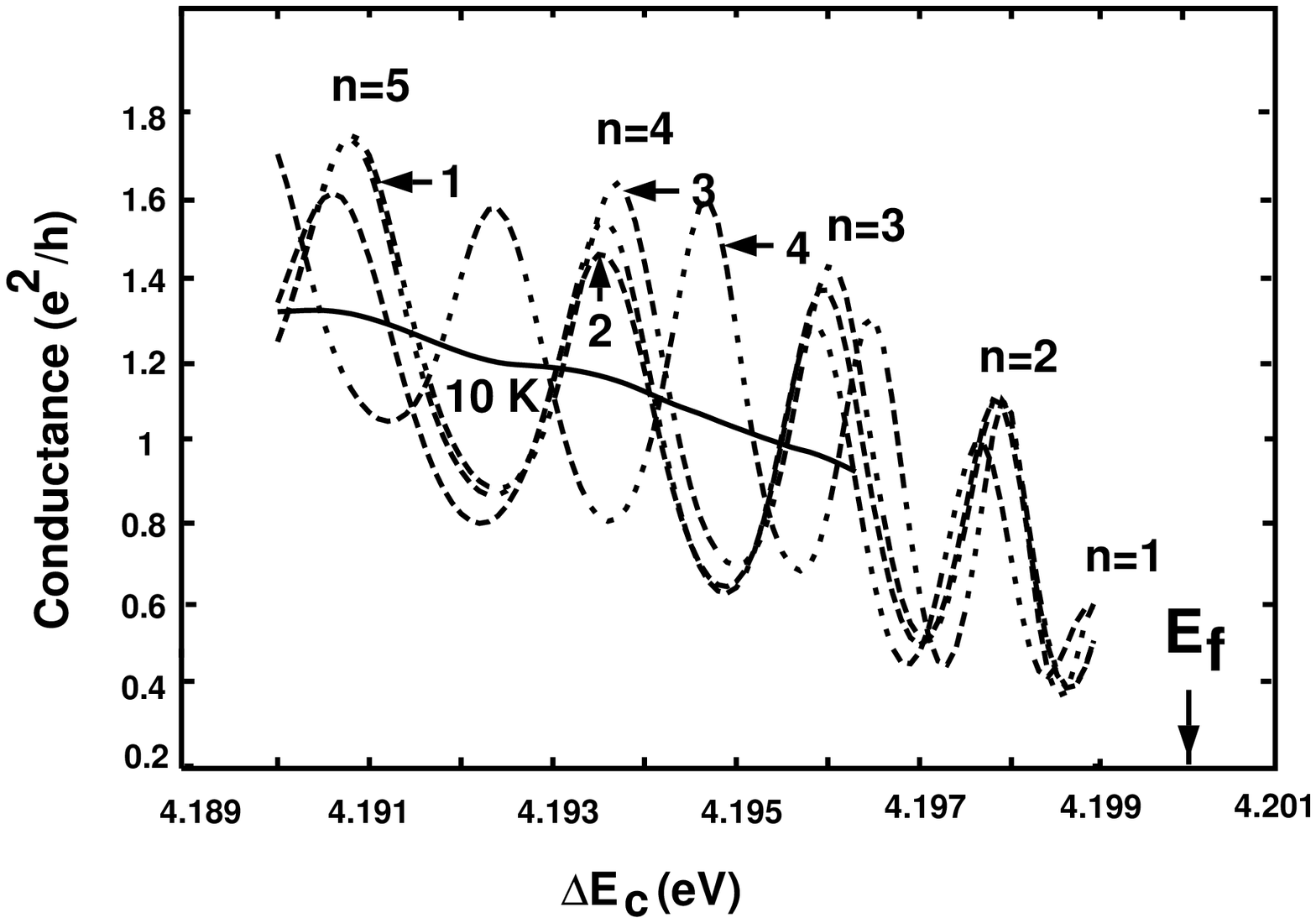,height=6.5in,width=6.5in}}
\end{figure}

\newpage
\begin{thebibliography}{10}

\bibitem{datta} 
S. Datta and B. Das, Appl. Phys. Lett., {\bf 56}, 665 (1990).

\bibitem{rashba}
E. I. Rashba, Sov. Phys. Semicond., {\bf 2}, 1109 (1960); 
Y. A. Bychkov and E. I. Rashba, J. Phys. C, {\bf 17}, 6039 (1984).

\bibitem{ohanian}
This is a textbook example of one-dimensional tunneling problems.
See for instance, H.C. Ohanian, {\it Principles of Quantum Mechanics}, 
Prentice Hall, New Jersey, p. 96 (1990).

\bibitem{moroz}
A.V. Moroz and C.H.W.Barnes, Phys. Rev. B, {\bf 60}, 14272 (1999); Phys. Rev. B 
{\bf 61}, R2464 
(2000).

\bibitem{mireles}
We use the same values as in F. Mireles and G. Kirczenow,
Europhys. Lett., {\bf 59}, 107 (2002).

\bibitem{nitta}
J. Nitta, T. Akazaki, H. Takayanagi, and T. Enoki,
Phys. Rev. Lett., {\bf  78}, 1335 (1997); Th. Sch\"apers, J. Nitta, H.B. 
Heersche, and 
H. Takayanagi,
Phys. Rev. B, {\bf 64}, 125314 (2001); Y. Sato, T. Kita, S. Gozu and S. Yamada,
J. Appl. Phys., {\bf 89}, 8017 (2001); Y. Sato, S. Gozu, T. Kita and S. Yamada, 
Physica 
E, {\bf 12}, 399 (2002).

\bibitem{schapers2}
Th. Sch\"apers, G. Engels, J. Lange, Th. Klocke, M. Hollfelder, and 
H. Luth, J. Appl. Phys., {\bf 83}, 4324 (1998).

\bibitem{numerical}
We compute the conductance using the Landauer formula by integrating
over an energy range from [ $E_f - 4 k_B T, E_f + 4 k_B T $]. For each 
temperature,
we limit the range of variation of $\Delta E_c$ so that both channels under the 
gate
are conducting for the range of energy considered. 

\bibitem{kriman}
A. M. Kriman and P. P. Ruden, Phys. Rev. B., {\bf 32}, 8013 (1985).

\bibitem{frohne}
R. Frohne and S. Datta, J. Appl. Phys., {\bf 64}, 4086 (1988).

\bibitem{grundler}
D. Grundler, Phys. Rev. B, {\bf 63}, 161307(R) (2001).

\bibitem{raichev}
O. E. Raichev and P. Debray,
Phys. Rev. B, {\bf 65}, 085319 (2002).

\bibitem{eirashba}
E. I. Rashba, Phys. Rev. B, {\bf 62}, 16267 (2000).

\bibitem{cahay}
S. Datta, M. Cahay and M. McLennan, Phys. Rev. B, {\bf 36}, 5655 (1987);
M. Cahay, M. McLennan and S. Datta, {\it ibid}, {\bf 37}, 10255 (1988).

\end{thebibliography}
\end{document}